\documentclass[11pt]{article}

\usepackage{amsmath, amssymb, wasysym, slashed, multirow,hyperref}
\usepackage{pdfpages}
\usepackage{cite}
\usepackage{times}

\newenvironment{sciabstract}{%
\begin{quote} }
{\end{quote}}

\newcounter{lastnote}

\title{Non-local Lagrangians from Renormalons and  Analyzable Functions}

\author
{Alessio Maiezza,$^{1\ast}$  Juan Carlos Vasquez$^{2\dagger}$\\
\\
\normalsize{$^{1}$Ruder Bo\v skovi\'c Institute, Bijeni\v cka cesta 54, 10000, Zagreb, Croatia,}\\
\normalsize{$^{2}$Universidad T\'ecnica Federico Santa Mar\'ia $\&$ CCTVal, Valpara\'iso, Chile}\\
\\
\small{ E-mail: amaiezza@irb.hr$^{\ast}$,juan.vasquezcar@usm.cl$^{\dagger}$}
}

\date{}

\begin{document}


\baselineskip16pt 


\maketitle


\begin{sciabstract}
We embed in a generalized Borel procedure the notion of renormalization and renormalons.
While there are several efforts in literature to have a semi-classical understanding of the renormalons, here we argue that
this is not the fundamental issue and show how to deal with the problem. We find that the effective Lagrangians describing the effects of renormalons are non-local in space but local in time. The quark-antiquark potential in QCD with an infinite number of fermions is also analyzed. The connection between the analyzable functions, the Callan-Symanzyk equation and the renormalons, provides an insight of a non-perturbative renormalization from the standard perturbative renormalization approach.
\end{sciabstract}

\section{Introduction}

Since the early days of the Quantum Field Theory (QFT) the problem of the infinities, appearing in the calculations of physical quantities and making the theory
apparently inconsistent, was noted by Pauli and Heisenberg~\cite{Heisenberg:1929xj}.
Fortunately, there is the well known procedure of renormalization that improve the convergence~\cite{Schwinger:1948iu,Schwinger:1948yk,Schwinger:1948yj,Schwinger:1949ra,Feynman:1948ur,Feynman:1948fi,Tomonaga:1946zz,Koba:1947rzy,Tomonaga:1948zz}. As a by product, the \emph{perturbative} renormalization  prescriptions~\cite{Dyson:1949bp} appear for processes that can be described using small coupling constants.
After renormalization, the original divergences disappear and the coupling constant expansion converges for infinitesimal couplings: all the infinities are reabsorbed in
the counterterms~\cite{Bogolyubov:1980nc}, which are a finite number of local operators of the same form of the renormalizable Lagrangian with arbitrary coefficients. Finally, requiring the truncated perturbative expansion to be invariant under arbitrary rescaling in the particular counterterms, leads to the renormalization group equations~\cite{Stueckelberg:1951gg,GellMann:1954fq}.

This would appear to be the whole story, but soon after, it was realized that after applying the perturbative renormalization prescription, the obtained series are still divergent~\cite{Dyson:1952tj} for finite, non-zero values of the coupling constant. They should be understood as asymptotic expansions for the expressions of any physical quantities. Then the fact that perturbation theory works is a possibility but it is not guaranteed \emph{a priori} from a purely mathematical point of view.

These divergent asymptotic series can be made convergent for finite values of the coupling constants by applying the Borel resummation procedure.
However,  this method may fail when the perturbative series behave as
$n!$, where $n$ is the order of the loop expansion because the Borel transform has singularities. When these singularities lie on the real and positive axis, the Laplace transform becomes ambiguous and the original series is not Borel resummable. In this case, one ends up in a situation in which the perturbation theory is not anymore able to capture the analytic structure of the Green's functions. This happens  at least for two known cases: instantons~\cite{Lipatov:1976rb}  and renormalons~\cite{tHooft:1977xjm}. The formers are regarded to be not so dangerous since asymptotic analysis can be used to treat them and the underlying physics is under control~\cite{tHooft:1977xjm}. Conversely, the physics associated with the renormalons is not well understood and this historically has been the obstacle to bypass them. Recently, a non-trivial generalization of the renormalon concept to the multi-coupling case has been developed in Ref.~\cite{Maiezza:2018pkk}.

There is still the possibility that the crux for circumventing the renormalons is not a specific physics or a semi-classical understanding of them\footnote{See Ref.~\cite{Argyres:2012vv} for a semiclassical interpretation of  the IR renormalons in a QFT in $\mathbb{R}^3\times S_1$.}, but instead a proper mathematical framework. This is the line that we follow in this work. We shall show that the proper mathematical environment to address the renormalon singularities is the framework of analyzable functions described by Costin in Ref.~\cite{Costin1995,costin1998,Borel-Ecalle}\footnote{Curiously,  in Ref.~\cite{Frishman:1979vh},  the necessity of a generalized mechanism is called for in order to consistently treat the infinite renormalon singularities.}. The generalization of the Borel resummation within the theory of analyzable functions allows to resum series that  cannot be unambiguously  resummed by the Borel procedure,  thus capturing  some genuine non-perturbative structures of the Green's functions.
The renormalons provide a specific realization of Wilson's renormalization~\cite{Wilson:1973jj}  from the perturbative renormalization theory. For scale invariant models, the new approach reveals the regions of universality in the multi-coupling parameter space.

The structure of this paper is organized as follows: in Sec.~\ref{parisi_discusion} we briefly discuss the notion of renormalons introduced by 't Hooft in Ref.~\cite{tHooft:1977xjm} for scalar field theories. Later in subsection~\ref{asParisi} and following Refs.~\cite{Parisi:1978bj,Parisi:1978iq}, we recall why an infinite number of higher dimensional, non-renormalizable operators appears in the Lagrangian when going beyond perturbation theory.
In Sec.~\ref{renormalons_phi4}  we apply the generalized resummation of Costin~\cite{Costin1995,costin1998,Borel-Ecalle} (and explained in more detail in App.~\ref{app1}) to resum the ultraviolet  renormalons in the $\phi^4$ model. We then show, in agreement with the Callan-Symanzyk equations (CS)~\cite{Callan:1970yg,Symanzik:1970rt}, that the resummed result could be described by the addition of a new piece proportional to single non-perturbative constant. We also briefly enumerate the theorems of Ref.\cite{Borel-Ecalle}  upon which the results found are based, in order to stress the robustness and to discuss the possible consequences.
In Sec.\ref{quarkantiquark_largeNf} we exploit the approach presented in Sec.~\ref{renormalons_phi4} for the quark-antiquark potential for a large number of fermions, showing that the renormalons capture the expected features for this model.
In Sec.~\ref{topological_considerations} we discuss the relationship between the Callan-Symanzyk equations,  renormalons, and scale invariance. Finally, in Sec.~\ref{conclusions} our conclusions are given.

\section{Perturbative renormalization and renormalons} \label{parisi_discusion}

\begin{figure}
 \centering
     \includegraphics[scale=0.45]{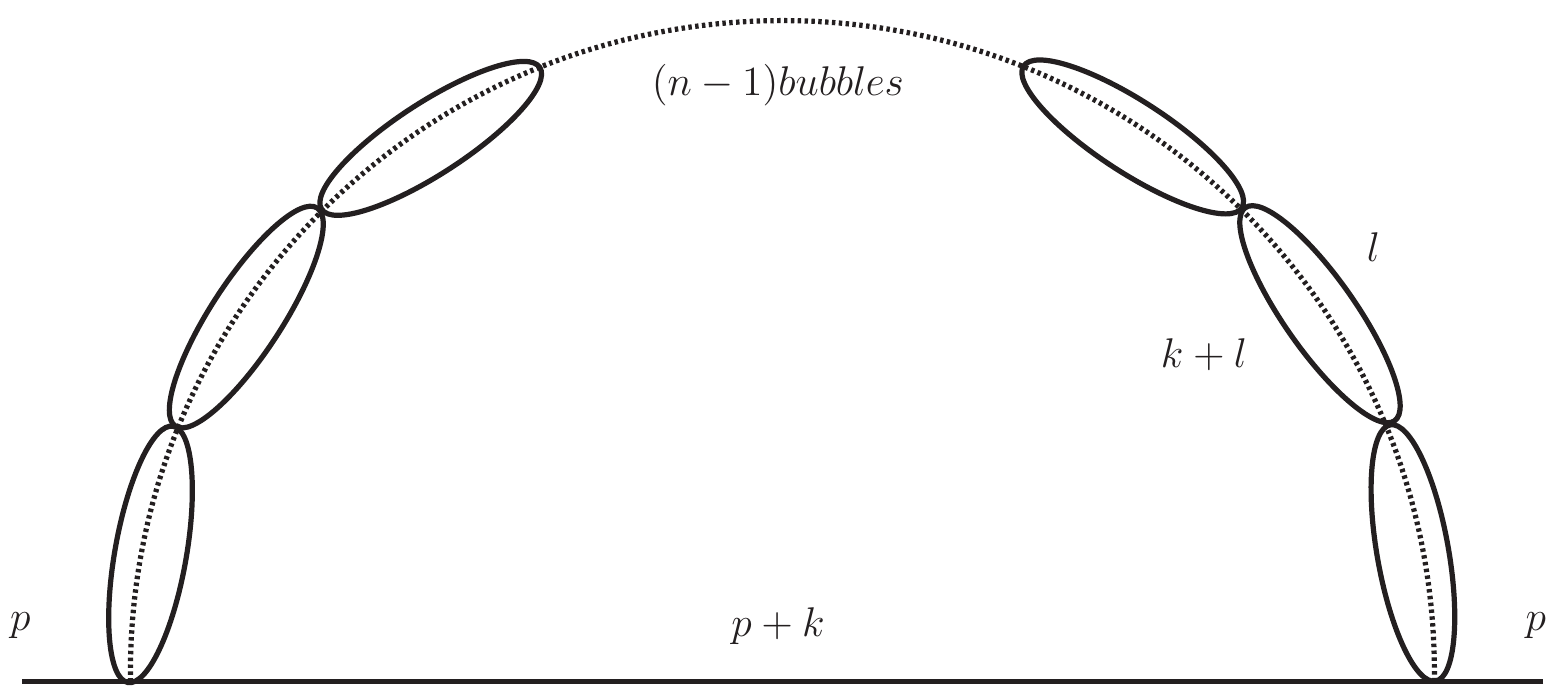}
     \includegraphics[scale=0.3]{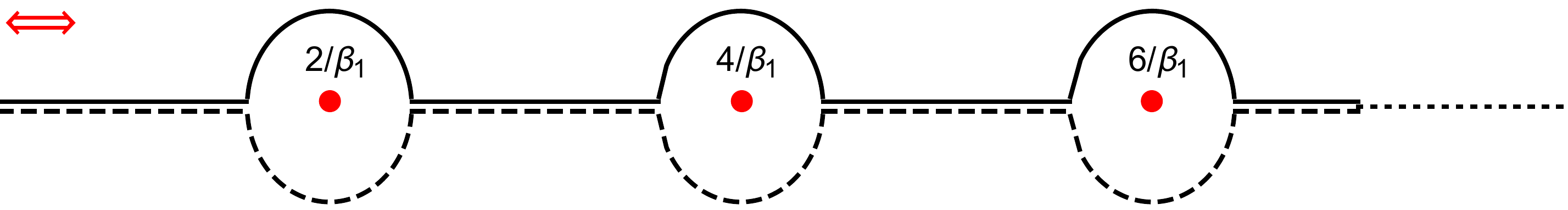}
  \caption{Pictorial illustration of the renormalons. Skeleton diagram for $\phi^4$ producing the renormalons (left panel);
   ambiguities related to the arbitrary choices of the contour modification in the Laplace transform (right panel).}
   \label{fig:poles}
\end{figure}

Let us do in this section a brief recap about the troubles of  the perturbative renormalization when pushed to the strong coupling regime. To this end,
consider the $\phi^4$ model with the renormalizable Lagrangian
\begin{equation}\label{basicL}
\mathcal{L}=\frac{1}{2}\partial_\mu \partial^\mu \phi -\frac{1}{2}m^2\phi^2- \frac{\lambda}{4!} \phi^4  \,.
\end{equation}
In the ultraviolet limit,  the Borel transform of the renormalized 2-point function, corresponding to a specific topology often referred as skeleton diagram~\cite{tHooft:1977xjm} and shown in Fig.~\ref{fig:poles}, contains singularities  (renormalons~\footnote{ For thermal and magnetic renormalons see Refs.~\cite{Loewe:1999kw,Correa:2019xvw} respectively. })  at,
\begin{equation} \label{renormalon_positions}
\text{Renormalons}: 2/\beta_1, 4/\beta_1, 6/\beta_1,... \,,
\end{equation}
being $\beta_1>0$ the  coefficient of the one-loop $\beta-$function,  $\beta(\lambda) = \mu d \lambda /d\mu= \beta_1 \lambda^2+\mathcal{O}(\lambda^4)$. It is worth recalling that the one-loop coefficient $\beta_1$  is just a constant of the problem that enters when calculating the  ``fish-diagrams'' chain that composes the skeleton diagram of Fig.\ref{fig:poles}.
For this simple model and after summing over all the diagrams with $n$ bubbles of the type shown in Fig.~\ref{fig:poles} , the poles of the Borel transform are described by the Hurwitz-Lerch function $\Phi$
\begin{equation}\label{ren}
\mathcal{B}(\Gamma^{(2)}_R(p)) \equiv \mathcal{G}^{(2)}_0(z)\propto -1/2 \beta_1 \Phi(-1,1,1-z\beta_1/2)+...= \sum _{i=1}^{\infty } \frac{(-1)^i}{2 i/\beta_1-z} + (\text{analytic terms})\,,
\end{equation}
 showing   an infinite number of simple poles on the real and positive axis spaced by a unit. The Borel transform of the skeleton diagram has infinite  singularities and their associated  infinite  ambiguities (seen pictorially in Fig.~\ref{fig:poles}) in the Laplace's transform, which should provide a finite output for the original (divergent) series. These ambiguities correspond to the arbitrariness in the regularization of the Laplace's integral that thus, signals the breakdown of the  perturbative approach, in particular of the perturbative renormalizability.

\subsection{Higher order operators}\label{asParisi}

We follow Parisi~\cite{Parisi:1978bj,Parisi:1978iq} and consider the renormalized,  1PI, $n$-point Green's function in momentum space obeying the Callan-Symanzyk  equation~\cite{Callan:1970yg,Symanzik:1970rt}
\begin{equation}\label{CS}
\left(-\sum_{i=1}^{n-1}p_i\cdot \frac{\partial }{\partial p_i} + \beta (\lambda)\frac{\partial}{\partial \lambda} + d_{n}-  n \gamma \right)\Gamma^{(n)}_R(p_1,p_2,...,p_{n-1})=0 \,,
\end{equation}
where $d_{n}= 4-n$ is the  canonical dimension of the $n$-point Green function.
Its solution can be  arranged at  the leading order as~\cite{tHooft:1977xjm,Parisi:1978iq}
\begin{equation}\label{On}
\Gamma^{( n)}_R\sim(\lambda)^{z_1}e^{(n-4)/(\beta_1\lambda)}f(p^2/\mu^2)\,,
\end{equation}
with $z_1=\frac{\beta _2
   (n-4)}{\beta _1^2}$, $f$ an arbitrary function of $p\equiv \sqrt{p_1^2+p_2^2+... }$ and $\mu$ a given renormalization scale. The Borel transform of Eq.~\eqref{On} yields
\begin{equation}\label{Higherpoles}
\mathcal{B}(\Gamma^{(n)}_R)=\mathcal{G}^{ (n)}_0(z)=\frac{\theta(z-\frac{n-4}{\beta_1})(z-\frac{n-4}{\beta_1})^{z_1-1}}{\Gamma(z_1)}\,,
\end{equation}
whose form is slightly different   from the one found in Ref.~\cite{Parisi:1978iq} but the position of the singularities are the same. However, here the only relevant point is that one gets singularities of the Borel transform
at $z=\frac{n-4}{\beta_1}$ for any Green function $\Gamma^{( n)}_R$ ($n>6$) in complete correspondence with the ones from the renormalons discussed above. The same conclusion has also been found in Ref.~\cite{Crutchfield:1979rt} using dimensional regularization.
Because of the B.H.P. theorem~\cite{bogoliubow1957,Hepp:1966eg} and according to Ref.~\cite{Parisi:1978iq}, the infinite ultraviolet   renormalons
ambiguities together with the Borel poles of $\Gamma^{( n)}_R$ can be then reabsorbed in a one-to-one correspondence into infinite higher-dimensional \emph{local} counterterms added to Eq.~\eqref{basicL}:
\begin{equation}\label{eff_lagrangian_renormalons}
\Delta\mathcal{L}(\phi) = \sum_{n\geq 6}^{\infty}c_{n} e^{-\frac{n-4}{\beta_1\lambda}}\phi(x)^{n}\,.
\end{equation}
The price to pay is to consider an infinite set of operators with  unknown (dimensionful) coefficients $c_{n}$ to the effective Lagrangian, thereby loosing predictivity. However, those higher operators are exponentially suppressed for small $\lambda$, and this is the crucial difference with a generic non-renormalizable lagrangian.
In the latter case, as emphasized in Ref.~\cite{Wilson:1970ag} and from a more generic point of view, the Lagrangian should include all the non(perturbatively)-renormalizable terms in the large coupling regime.

With a different approach based on the resurgence and the analyzable functions, we shall show that only one non-perturbative constant suffices to unambiguously describe the effect of renormalons.

\section{Resummation of the renormalons in the  $\phi^4$ model }\label{renormalons_phi4}

A resummation procedure is an isomorphism close under all operations,  which maps a divergent formal series to a unique function.
The standard Borel resummation is an isomorphism and commutes with the algebraic operations,  integration, and differentiation. It allows one to resum many divergent series. As shown above, this is not the case for  renormalons. The central issue of this section is to stress that one could consider a generalized isomorphism for resurgent functions able to go beyond the Borel resummation, instead of approaching the problem as in subsection~\ref{asParisi}. Indeed, such a generalized procedure does exist and it has been developed in Refs.~\cite{Costin1995,costin1998,Borel-Ecalle}.

The formalism exposed in Refs.~\cite{Costin1995,costin1998,Borel-Ecalle} can be applied to QFT. This formalism may be understood in a twofold way, namely:  given an ordinary differential equation (ODE) one can find its transseries solution and, from it, infer the analytic structure of the solution. This procedure is called the ``analysis"  of the function satisfying the ODE under consideration. The other aspect is that, given a  formal divergent series (or transseries) with poles in the positive Borel axis, one may use the generalized Borel resummation procedure of Refs.~\cite{Costin1995,costin1998,Borel-Ecalle}, and this reconstruction is referred to as the ``synthesis"  of the trasseries  one is interested in. The mathematical robustness of the resummation algorithm is guaranteed within the specific problem of ODE, which is the case for the renormalons, as we shall discuss.

Consider for example the contribution to the Borel transform $\mathcal{G}^{(2)}_0(z)$ of the two-point Green's function $\Gamma_R^{(2)}(p)$ shown in Eq.~\eqref{ren}. The generalized resurgence formalism connects $\mathcal{G}^{(2)}_0(z) $ to  the non-perturbative functions
$\mathcal{G}^{(2)}_k(z) $ with $k\geq1$. In fact, $\mathcal{G}^{(2)}_0(z) $ has an infinite number of simple poles in the positive  real line (renormalons),   such lines  are called Stoke lines, which in our case is just one,  as illustrated in Fig.~\ref{fig:poles}. Applying the Generalized Borel resummation of App.\ref{app1} to Eq.~\eqref{ren}, the non-perturbative expression for $\Gamma_R^{(2)}(p)$  is given by
\begin{align}\label{final}
\Gamma_R^{(2)}(\lambda)&\sim  \int_0^\infty  \left(\mathcal{G}^{(2)}_0\right)^{bal}  e^{-\frac{z}{\lambda}} dz +Ce^{-\frac{2}{\beta_1\lambda}}\int_0^\infty  \left(\mathcal{G}^{(2)}_1\right)^{bal} e^{-\frac{z}{\lambda}} dz+i\pi^2m^2 \lambda\log2 \nonumber \\
&=-(i\pi^2m^2) \left[PV\left(  \int_0^\infty \Phi(-1,1,1-\frac{\beta_1z}{2}) e^{-\frac{z}{\lambda}} dz \right) -\lambda\log2+\frac{4 \pi^2 C}{S} \left(\frac{e^{-\frac{2}{\beta_1\lambda}}}{1+e^{-\frac{2}{\beta_1\lambda}}}\right)\right]\,,\nonumber \\
&\simeq-(i\pi^2m^2) \left[PV\left(  \int_0^\infty \Phi(-1,1,1-\frac{\beta_1z}{2}) e^{-\frac{z}{\lambda}} dz \right) -\lambda\log2+\frac{4 \pi^2 C}{S}\frac{E^2}{\Lambda^2+E^2}    \right]\,\nonumber\\
\end{align}
where in the last step we have explicitly introduced the non-perturbative scale $\Lambda$ at which the piece proportional to $C/S$ starts to be quantitatively relevant  (see the discussion below Eq.~\eqref{nonlocal_CT} for a detailed interpretation of this scale), $e^{-\frac{2}{\beta_1\lambda}}\propto E^2/\Lambda^2$ with $E$ being the energy.
$PV$ denotes the Cauchy principal value and it is given by
\begin{equation}
PV\left(  \int_0^\infty \Phi(-1,1,1-\frac{\beta_1z}{2}) e^{-\frac{z}{\lambda}} dz \right) =\sum_{n=1}^{\infty} (-1)^n  e^{-\frac{2 n}{\beta_1 \lambda}}Ei(\frac{2 n}{\beta_1 \lambda})\,,\label{PV}
\end{equation}
being $Ei(x)$ the exponential integral function.
The term $\lambda \log2$ has been introduced for convenience and is such that it cancels the linear term in $\lambda$ in the asymptotic expansion of the principal value part.  The Eq.~\eqref{final} is a non-perturbative expression and only the asymptotic  (divergent) expansion of the part with the principal value has an interpretation in terms of Feynman loops. Furthermore, $PV$ provides a complete description of models with coupling constants smaller than the first renormalon limit. Beyond this limit, the non-perturbative sectors (associated with the $ \mathcal{G}^{(2)}_i$ for $i\geq1$) need to be included in the description. In Sec.~\ref{topological_considerations}, we further discuss this issue in more detail in connection with scale invariance. Since the renormalon bubble chain may be inserted in any n-point Green's function, this constant $C/S$ should be unique in a given model.

\begin{figure}
 \centering
     \includegraphics[scale=0.4]{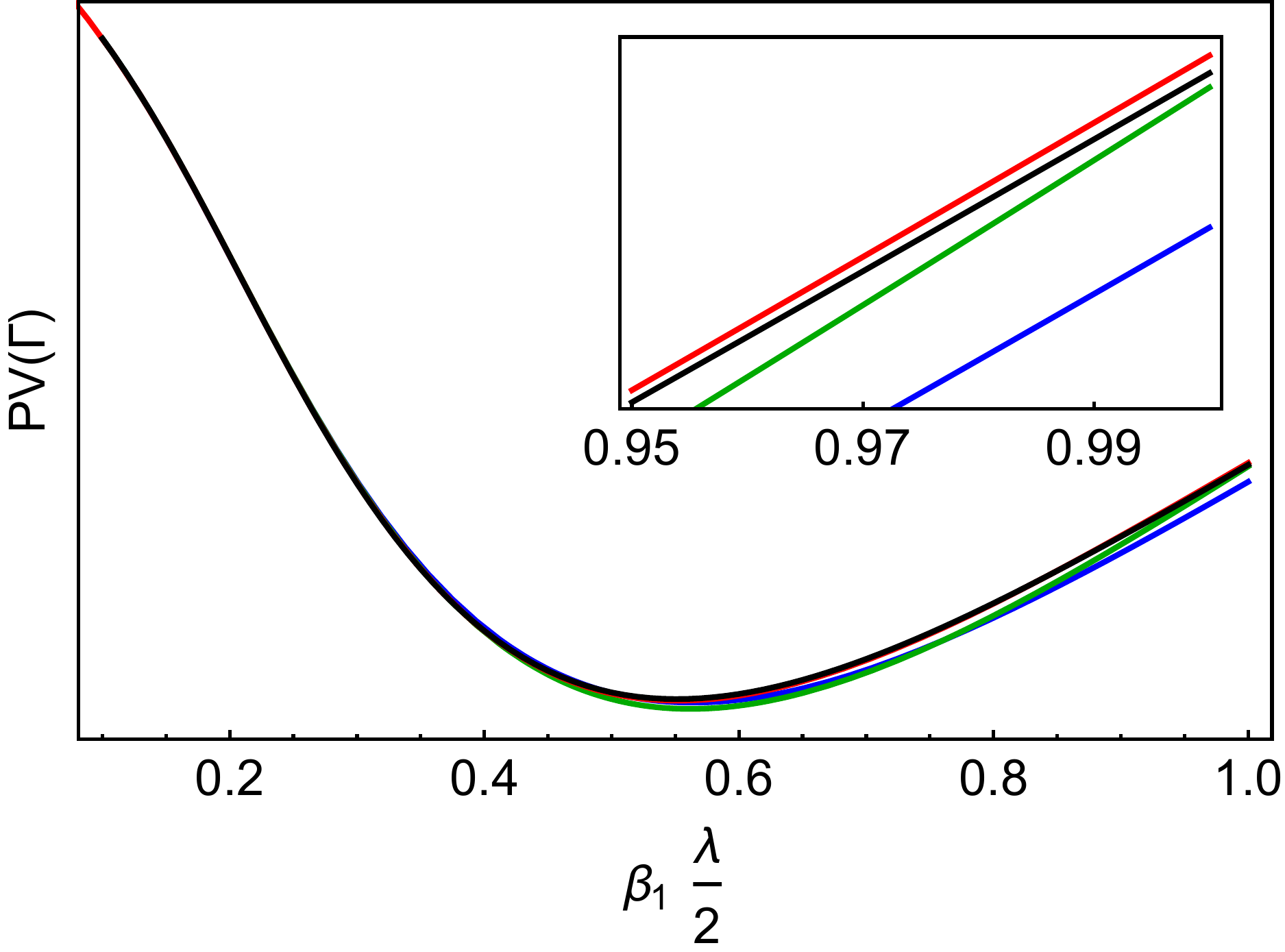}
  \caption{Comparison between the exact numerical result of the principal value part of Eq.~\eqref{final} (black line) and the first few Meijer-G-approximants (color lines) of the asymptotic expression for the principal value part in Eq.~\eqref{final}.  Blue, green, and red represent the first, second and third Meijer-G-approximants. For convenience, we have plotted as a function of the variable $\beta_1\lambda/2$.}
   \label{fig:mejierG}
\end{figure}

It is interesting to notice that the algorithm proposed in Ref.~\cite{Mera:2018qte} is able to capture from the perturbative expansion precisely the regularized principal value function shown in Eq.~\eqref{PV}. In Fig.~\ref{fig:mejierG}, we show the result of the numerical computation of the principal value part shown in Eq.~\eqref{PV} and show that the real part of the first Meijer-G-approximants coincides with the exact numerical values with good accuracy. This is in agreement with the conjecture in Ref~\cite{Antipin:2018asc}.

\paragraph{Synthesis of renormalons.}

Let us elaborate in more detail the result shown in Eq.\eqref{final}: the non-perturbative function $\mathcal{G}^{(2)}_1(z)$ can be obtained by applying the isomorphism explained in App.~\ref{app1} and it is given by,
\begin{equation}
S \mathcal{G}^{(2)}_1 = \left(( \mathcal{G}^{(2)}_0)^- -(\mathcal{G}^{(2)}_0)^+ \right)\circ \tau_1 \propto -2\pi i \sum_{p=1}^{\infty} (-1)^p \delta[z+(1-p)\frac{2}{\beta_1}]\,,
\end{equation}
where we have used  Eqs.~\eqref{eq:resurgence}-\eqref{eq:analyticcontinuation}. The next step is to find the expression for the higher $\mathcal{G}^{(2)}_i$. For $\mathcal{G}^{(2)}_2$ we have,
\begin{equation}
S^2 \mathcal{G}^{(2)}_2 = \left(( \mathcal{G}^{(2)}_0)^- -(\mathcal{G}^{(2)}_0)^{-1+} \right)\circ \tau_2=0\,,
\end{equation}
and in the same way $\mathcal{G}^{(2)}_i= 0 $ for $i\geq3$. This simple result is a consequence of the simple pole structure of the Hurwitz-Lerch function.

Once the expressions for the $\mathcal{G}^{(2)}_i$ are known, the \emph{balanced average} for them needs to be computed. For $\mathcal{G}^{(2)}_0$ and $\mathcal{G}^{(2)}_1$ we have:
\begin{align}\label{G1}
& \left(\mathcal{G}^{(2)}_0\right)^{bal} = \frac{1}{2}\left[  (\mathcal{G}^{(2)}_0)^+ +(\mathcal{G}^{(2)}_0)^- \right],\quad  \left(\mathcal{G}^{(2)}_1\right)^{bal} = \frac{1}{2}\left[  (\mathcal{G}^{(2)}_1)^+ +(\mathcal{G}^{(2)}_1)^- \right]\,,
\end{align}
as shown in Eq.~\eqref{final}. It can be seen from Eq.~\eqref{balanced:average} that the balanced average is  a potentially complicated operation. In our case the simplification is due to the fact that only $\mathcal{G}^{(2)}_0$ and $\mathcal{G}^{(2)}_1$ are non-zero. The last step is to perform the Laplace transform of Eq.~\eqref{G1}. While the $PV$ in Eq.~\ref{final} comes from the $\mathcal{G}_0^{(2)}$ part, the non-perturbative piece comes from
\begin{equation}
\int_0^\infty \mathcal{G}^{(2)}_1(z) e^{-\frac{z}{\lambda}} dz \propto-\frac{2\pi i}{S_0} \sum_{p=1}^{\infty} (-1)^pe^{-\frac{p-1}{\lambda}\frac{2}{\beta_1}}\,.
\end{equation}
It is worth to emphasize that this resummation procedure  starts with perturbation theory, which the only required input of the  formalism developed in Ref.~\cite{costin1998}. Unlike the usual perturbative approach, it allows one to go to the large coupling regime in a well-defined manner, at least for the renormalons. It is not known whether this formalism may be able to resum other possible unknown ``bad'' series that leads to an $n!$ divergence in the perturbative expansion.  In any case, it is reasonable to assume that these new bad series do not modify the position of the renormalon poles~\cite{Altarelli:1995kz} and hence that the conclusions  here presented are still  valid in the more general case.

\paragraph{Remarks.}
One should stress that the mathematical robustness of our finding is built upon  \emph{theorems}:  theorem 5.67 (analytic properties),  proposition  5.77 (median average), proposition 5.113 (recursive relations) and theorem 5.120 (applicability) of Ref.~\cite{Borel-Ecalle}.

We conclude that the generalized Borel resummation strongly alleviates the problems of renormalons. Although the transseries parameter (see above $C/S$) is not calculable, in contrast to what can be done for the instantons\footnote{For comprehensive reviews on resurgence and transseries for instantons see Refs.~\cite{Aniceto:2018bis,Dorigoni:2014hea}.}~\cite{Lipatov:1976rb}, the arbitrariness is contained in a single constant. The difference among the instantons and renormalons is known since the seminal work of 't Hooft, namely one cannot implement the semiclassical method for the latter.

It is also instructive to find the effective Lagrangian that produces the correction proportional to $C/S$. The Fourier transform of Eq.~\eqref{final} gives
\begin{equation}
\Gamma^{(2)}_R(x_1,x_2) \sim \mathcal{K}m^2\delta(t_1-t_2) \left(  \delta^{(3)} ( \vec{x}-\vec{y}    )-\frac{ \Lambda ^2e^{-\sqrt{\Lambda^2+m^2}\, |\vec{x}-\vec{y}    |}
   }{8\pi |\vec{x}-\vec{y}    |}\right)\,,
\end{equation}
where $\mathcal{K}= -4\pi(i\pi^2)\left(\text{4 }\pi ^2\frac{C}{S}\right)2\sqrt{2} \pi ^{3/2}$. Then,  the term proportional to $C/S$ in Eq.~\eqref{final}, is described by the addition of a  countertem in the Lagrangian \emph{local} in time, but  \emph{non-local}  in space  of the form
\begin{equation}\label{nonlocal_CT}
\Delta\mathcal{L}(x)=-  \mathcal{K}m^2  \Lambda^2 \int d^3\vec{y}\,\,\,  \frac{e^{-\sqrt{\Lambda^2+m^2}\, |\vec{x}-\vec{y}    |}}{8\pi |\vec{x}-\vec{y}    |}  \phi(t,\vec{x})\phi(t,\vec{y}) \,.
\end{equation}
Note that a formally similar equation has been proposed in the  description of  perturbative confinement in Ref.~\cite{tHooft:2003lzk}, in which the confinement itself is directly built-in the QCD Lagrangian by a judicious choice of the counterterms. In our case the interpretation is a bit different: Eq.~\eqref{final} emerges from a self-consistent calculation and only afterwards  it is reviewed as a non-local counter-term. The meaning of $\Lambda$ is now probably clearer, indeed it signals the scale at which the effective Lagrangian in Eq.~\eqref{nonlocal_CT} must be included in the description. As already emphasized in Ref.~\cite{tHooft:2003lzk}, the kind of acceptable counterterms  that produce unitary $S$-matrix at all stages, may be non-local in space but  local in time, which is precisely the result that we obtain in Eq.~\eqref{nonlocal_CT}. At first glance, the non-perturbative scale $\Lambda$ resembles the usual Landau pole scale. Conversely, we  are pointing  out that the Laundau pole  is rather a symptom in the perturbative approach  of the necessity of including  non-local counterterms of  the type shown in Eq.~\eqref{nonlocal_CT}, because when the coupling is large enough the contribution proportional to $C/S$ in Eq.~\eqref{final} becomes relevant. Since this new Lagrangian is not the original renormalizable one in Eq.~\eqref{basicL}, no Landau pole emerges.

A final comment is in order: in Ref.~\cite{tHooft:1977xjm}, it was argued that Green's function has singularities at finite values of the coupling constants at $1/\lambda = real- \frac{\beta_1}{2}(2n+1)\pi i$, where the real number may be arbitrarily large. When the   renormalons are unambiguously resummed,  the   Eq.~\eqref{final} converges for $\Re e(\lambda) < \lambda_0:\lambda_0\in \mathbb{R}$, according to theorems 5.120 and 5.125 of Ref.~\cite{Borel-Ecalle}  and in agreement with Ref.~\cite{tHooft:1977xjm}. For non-asymptotically free models, this is a substantial improvement with respect to the usual perturbation theory, where according to the state-of-the-art knowledge, the computed expressions are strictly valid for infinitesimal couplings and therefore mathematically ill-defined for any finite value of the coupling constant.


\paragraph{Final topological considerations.}

Following again Refs.~\cite{costin1998,Borel-Ecalle}, it is interesting to give a topological interpretation of the non-perturbative expression in Eq.~\eqref{final}.  To this end, consider the homotopy classes of curves in the complex Borel plain,  starting at the origin and crossing the Stokes line at most once: $(\mathcal{G}^{(2)}_0)^{\pm}$   are the analytic continuations along the curves that start at the origin and go above and below the Stokes line, without crossing it, respectively; $(\mathcal{G}^{(2)}_0)^{-1+}$ is the analytic continuation along a curve starting at zero from below,
crossing the Stokes line between $2/\beta_1$ and $4/\beta_1$, and going above it. These analytic continuations are connected with the origin, and hence they are related to perturbation theory. The Eq.~\eqref{eq:resurgence} involves all these analytic continuations and in particular, the $\tau$ operator (see  App.~\ref{app1}) is responsible for the disconnection from perturbation theory.  Specifically,  in Eq.~\eqref{final} the disconnected object is $\mathcal{G}_1^{(2)}$ and is, for instance,   the origin of the piece proportional to $C/S$. This is a topological reason why the perturbative proofs of renormalizability for  the $\phi^4$ model give no hint that the theory is actually non-renormalizable for any non-zero value of the coupling $\lambda$.

\section{The quark-antiquark potential for an  infinite number of fermions }\label{quarkantiquark_largeNf}

 \begin{figure}\centering
     \includegraphics[scale=0.35]{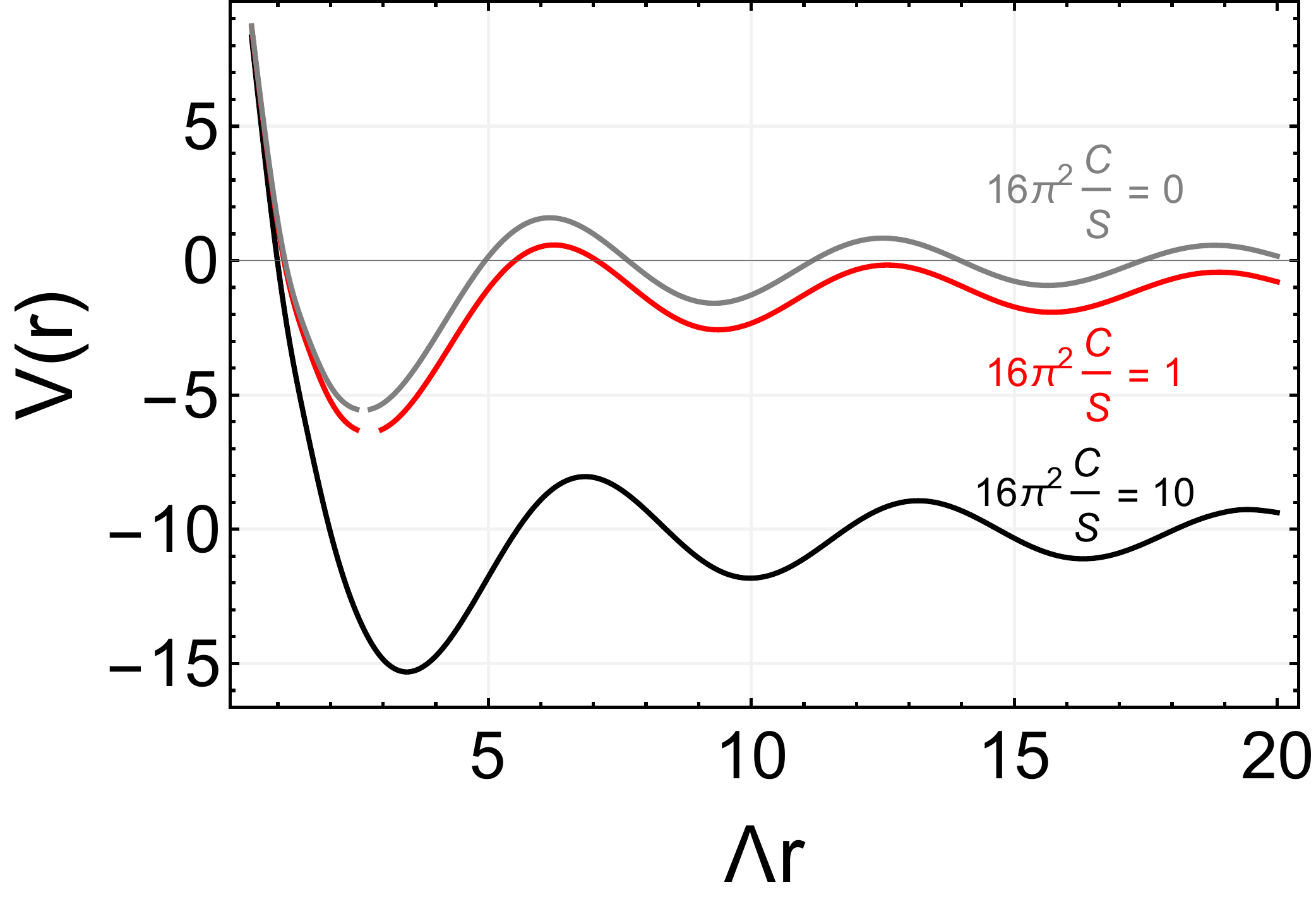}
  \caption{ Static potential between  quarks in $SU(3)$ in the limit of large number of fermion flavors $N_f$. }
  \label{fig:pot}
\end{figure}

We show in a simplified model, how the expected behavior of the quark-antiquark potential can be derived from the renormalizable Yang-Mills Lagrangian.
To this end consider $SU(3)$ gauge theory with a large number of fermions $N_f$.  The quark anti-quark potential for this model was calculated in Ref.~\cite{Aglietti:1995tg}, where  they found for the Borel transform of the potential $V ( r ) = \int _ { 0 } ^ { \infty } \mathrm { d } z \exp \left[ - z/ \left( b _ { 0 } \alpha \right) \right] \tilde { V } ( r )$,  the following  result:
\begin{equation}\label{borel_pot}
\tilde{V}(r,z) = - \frac { 4 \mathrm { e} ^ { - c z } } { b _ { 0 } } \frac { 1 } { r } ( \mu r ) ^ { 2 z } \frac { \Gamma \left( \frac { 1 } { 2 } + z \right) \Gamma \left( \frac { 1 } { 2 } - z \right) } { \Gamma ( 2 z + 1 ) }\,,
\end{equation}
$c$ is a scheme dependent constant and is equal to zero  in the momentum subtraction scheme and $\beta ( \alpha ) = \mu ^ { 2 } \partial \alpha / \partial \mu ^ { 2 } = - b _ { 0 } \alpha ^ { 2 } + \mathcal { O } \left( \alpha ^ { 3 } \right)$.

The function shown in Eq.~\ref{borel_pot}  has an infinite number of singularities at $z_{pole}= k+\frac{1}{2}$, where $k\geq0$. Hence,  we apply the formal resummation procedure explained in App.\ref{app1} in order to get the ambiguity free expression for the quark anti-quark potential, which is
\begin{equation}\label{ecalle_pot}
\sigma\left[V(r)\right] \mapsto - \frac{1}{r}\frac{4}{b_0} PV\left[ \int_0^{\infty} dz \left(  \Lambda r \right)^{2z} \frac { \Gamma \left( \frac { 1 } { 2 } + z \right) \Gamma \left( \frac { 1 } { 2 } - z \right) } { \Gamma ( 2 z + 1 ) }
    \right] + \frac{16\pi^2   C}{S} \left[\frac{\Lambda r -\sin\left(\Lambda r\right)}{b_0 r} \right]\,,
\end{equation}
where $\Lambda^2/\mu^2 = e^{-\frac{1}{b_0\alpha[\mu^2]}}$ and $C,S$ are non-perturbative arbitrary constants. Let us recall that the link between the appearance of renormalon ambiguities and an energy scale is generic and thus holds even in the real QCD, as clearly emphasized recently in Ref.~\cite{Cvetic:2018qxs}.
In Fig.~\ref{fig:pot}, we show the static potential found in Eq.~\ref{ecalle_pot} for several values of $\frac{16\pi^2 C}{S}$.  As evident from the figure, the necessary energy to put two quark at zero distance is infinite. This is the ultraviolet version of the IR confinement in real QCD. On the other hand, the quarks behave as free particles at large distances. This is a positive indication that the infrared and ultraviolet confinement properties are interchanged with respect to real QCD.
This analysis shows that the expected behavior of the quark anti-quark potential for the $SU(3)$ gauge theory (in the large $N_f$ limit) may be correctly described when resumming the renormalons.

It is worth stressing that, as far as renormalons are concerned,  there is no fundamental difference between this example and the real world QCD. The practical limitations of real QCD are that on one side, $\tilde{V}(r,z)$  is computed up to a finite number of loops (three loops \cite{Peter:1996ig}), and on the other side, instanton's singularities are also present. Notice that these two issues are absent in the large $N_f$ expansion, since the Borel transform of the static potential is known and instanton's singularities go away in the large $N_f$ limit~\cite{PhysRevD.15.1655,KOPLIK1977109}. It is clear that expanding around the absolute minimum  the potential could be approximated as a sum of a Coulomb part plus a linear potential, and this is in agreement with  the analytic expression for the static potential used and derived in Refs.~\cite{tHooft:2003lzk,Sumino:2003vk} respectively.

Notice that for pure Yang-Mills models, there are poles in the positive real axis of the  Borel transform from the contribution of zero momenta, which are the infrared renormalons.  From the point of view of the formalism presented in Ref.~\cite{costin1998,Borel-Ecalle}, the resummation of these singularities poses no difficulties.   Remarkably,  the counterterm in Eq.~\eqref{nonlocal_CT}  is of the same  type of the one proposed in Ref.~\cite{tHooft:2003lzk} describing the permanent confinement of electric charges. Hence,  we would expect that by ressuming the infrared renormalons with the formalism here used,  these kind of terms would naturally appear in the effective Lagrangian of QCD at low energies.  Further calculations along these lines are left for a future work.

\section{The Callan-Symanzyk equation, renormalons and  scale invariant theories}\label{topological_considerations}

 \begin{figure}\centering
     \includegraphics[scale=0.37]{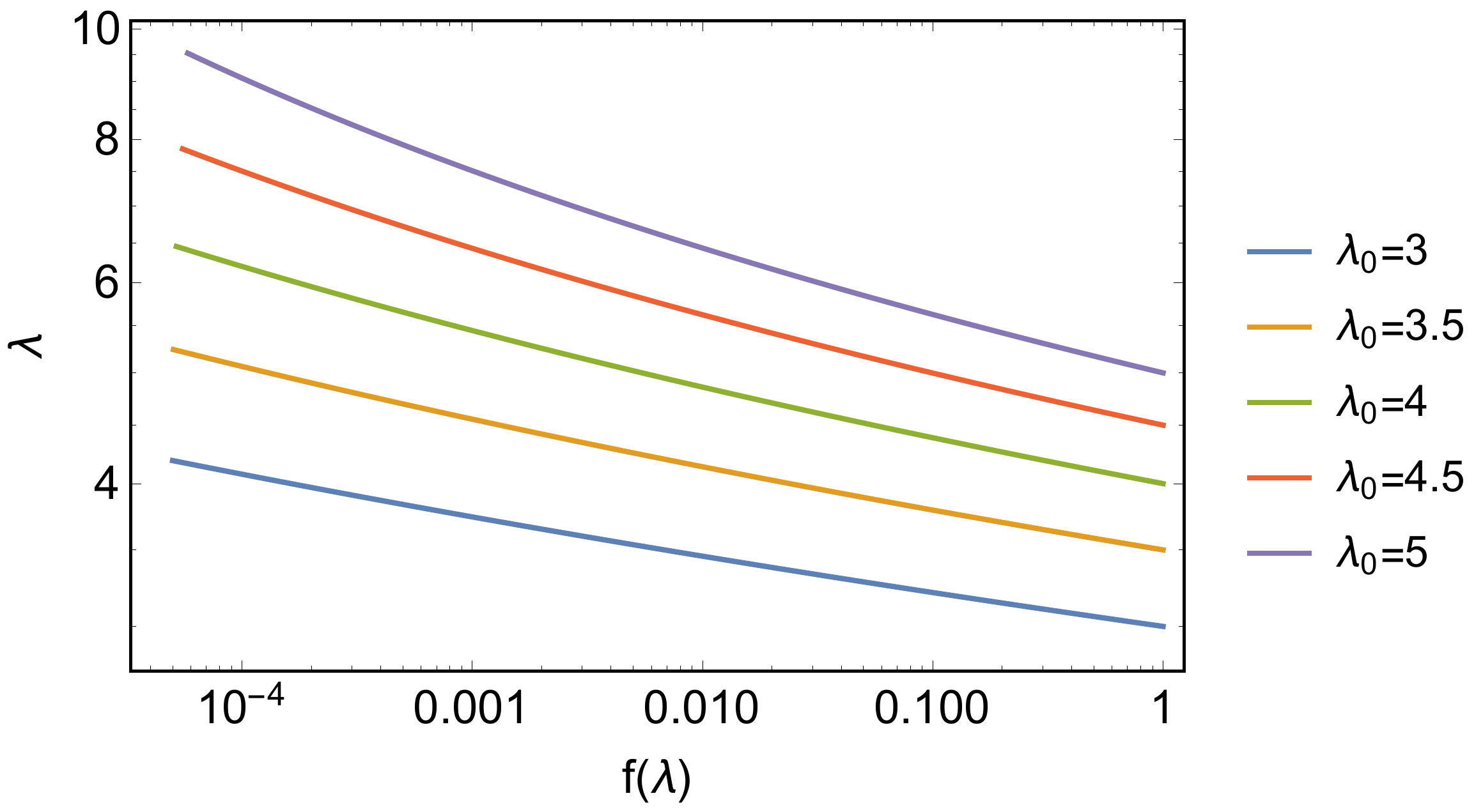}
  \caption{Evaluation of the function $f(\lambda)$ in Eq.~\eqref{match} via the Meijer G-function algorithm~\cite{Mera:2018qte}, as in Ref.~\cite{Antipin:2018asc}.}
  \label{fig:match}
\end{figure}

\subsection{Renormalons and the Callan-Symanzyk equation.}

The skeleton diagram in Fig.~\ref{fig:poles} is a particular topology of a two-point Green's function $\Gamma_R^{(2)}$ that must obey  the CS equation, which  is the Ward's identity of  the anomalous scale invariance and thus valid beyond perturbation theory. The general solution to the CS equations for the two-point function has the general solution
\begin{equation}\label{asPeskin}
\Gamma_R^{(2)}(p\rightarrow 0, \lambda(\mu))=-im^2g(\lambda_0(\lambda))\exp\left[\int_{\lambda_0}^{\lambda(\mu)}  \frac{2(\gamma(\lambda')  - 1)}{\beta(\lambda')}d\lambda'\right]\,,
\end{equation}
where $g(\lambda_0)$ is an arbitrary function that cannot be determined a priori from general principles and $\lambda_0$ is the coupling evaluated at the initial renormalization scale $\mu_0$. Notice the dependence of $\lambda_0$ in  $\lambda$ through the  renormalization group running. In perturbation theory,  one calculates $\Gamma^{(2)}_R$
in powers of $\lambda$ and fixes $g(\lambda_0(\lambda))$ by matching the calculation with the expansion of Eq.~\eqref{asPeskin}  (see for example Ref.~\cite{Peskin:1995ev}). For the renormalons,  one has to follow the same logic, except that one does not perform the perturbative expansion in Eq~\eqref{asPeskin}. Since  $g(\lambda_0(\lambda))$  is a priori an arbitrary function, one  must therefore  match the Eq.~\eqref{final} with Eq.~\eqref{asPeskin} in order to find this unknown function. Exactly the same logic applies for all the higher $n$-point Green's functions.  This non-perturbative matching,  must be done with the non-perturbative expressions of the $\beta(\lambda)$ and $\gamma(\lambda)$ functions. Expressions that may be obtained, for instance,  by using the Borel-hypergeometric resummation of Refs.~\cite{Mera:2014sfa,Mera:2018qte}, but applied to the $\beta$ and $\gamma$ functions in the spirit of Ref.~\cite{Antipin:2018asc}. Explicitly, this matching reads as
\begin{equation}\label{match}
g(\lambda)=\frac{\pi^2}{f(\lambda)}\left[PV\left(  \int_0^\infty \Phi(-1,1,1-\frac{\beta_1z}{2}) e^{-\frac{z}{\lambda(\mu)}} dz \right) -\lambda\log2+\frac{4 \pi^2 C}{S} \left(\frac{e^{-\frac{2}{\beta_1\lambda}}}{1+e^{-\frac{2}{\beta_1\lambda}}}\right)\right]\,,
\end{equation}
where $f(\lambda) = \exp\left[-\int_{\lambda_0}^{\lambda(\mu)}  \frac{2(\gamma(\lambda')  - 1)}{\beta(\lambda')}d\lambda'\right]$ is numerically calculated in Fig.~\ref{fig:match} for different  non-perturbative values of $\lambda_0$.

What are the differences with  respect to the perturbative procedure quoted in subsection~\ref{asParisi}? In that approach,  one does not get Borel singularities
for $\Gamma_R^{(2)}$ in the positive real axis,  but does  get instead for $\Gamma_R^{(2 n)}$ with $n\geq3$. However,  $\Gamma_R^{(2)}$ manifestly shows Borel singularities from the direct computation of the skeleton diagram, meaning indeed,  that the results are beyond perturbation theory. Conversely, in subsection~\ref{asParisi}, the possibility to reabsorb the non-perturbative ambiguities in $\phi^n$ is shown to be true in perturbation theory and at the same time assumed to be valid beyond the perturbative regime. It seems that this is a quite dared hypothesis. The issue may be  also understood  from a more direct point of view:
specifically, it results from Eq.~\eqref{Higherpoles} that the Borel transform of $\mathcal{G}^{(6)}_0$ has singularities of the form
\begin{equation}\label{poles_list}
\text{poles: } (z-\frac{n-4}{\beta_1})^{-1+z_1}\,,
\end{equation}
with $z_1$ evaluated via perturbative expansion as in subsection~\ref{asParisi}. However, when $\lambda$ is considered in a non-perturbative regime, the analytic structure might be completely different from the one suggested in Eq.~\eqref{poles_list}. Notice also that for small $\lambda$, $g(\lambda)= 1 +\mathcal{O}(\lambda)$ and in this case performing  the Borel transform of the solution in Eq.~\eqref{asPeskin} is equivalent to the Borel transform of the expanded CS equations, as done in Refs.~\cite{Parisi:1978bj,Parisi:1978iq}.

Nevertheless, we argue  that the results quoted in subsection~\ref{asParisi} may hold even beyond perturbation theory. Doing OPE~\cite{PhysRev.179.1499} to Eq.~\eqref{nonlocal_CT}, still one ends up with an expression in the form Eq.~\eqref{eff_lagrangian_renormalons}. Thus OPE may be seen as a non-perturbative bridge among Eq.~\eqref{final} and the approach in subsection~\ref{asParisi} (see Refs.~\cite{Beneke:1994sw,Bigi:1994em,Beneke:1998ui,Broadhurst:2000yc,Shifman:2013uka}  for  applications in QCD of the OPE and renormalons).

A conclusive comment is in order: the resummation of Ref.~\cite{costin1998} has been definitively proven to be an isomorphism within the specific problem of ODE, and this is consistent with Eq.~\eqref{final}, whose ODE is indeed the CS equation.

\subsection{Renormalons and scale invariant theories}

In this subsection our purpose is to clarify the relationship between the renormalons and scale-invariant theories. Consider a non-asymptotically free theory with coupling constant $g$ that exhibits a non-Gaussian  fixed point at $g=g_*$. When $g$ approaches to $g_*$, one can expand $\beta(g) = \beta'(g_*) (g-g_*) +\beta''(g_*)(g-g_*)^2+\mathcal{O}((g-g_*)^3)$ and $\gamma(g)= \gamma'(g_*)(g-g_*)+\mathcal{O}((g-g_*)^2)$, where the prime  denotes derivative with respect to the coupling constant. In this case the CS equation takes the form
\begin{align}\label{ASI1}
\frac{d \Gamma^{(n)}_R }{d g } =&  \left[ - \frac{4-n}{\beta'(g_*) }\frac{1}{g-g_*}+ \left(\frac{\gamma' n}{\beta'}-\frac{(n-4) \beta''}{\beta'^2}\right) +\mathcal{O}(g-g_*)  \right] \Gamma^{(n)}_R  \nonumber \\
& \equiv \left[ \frac{d_0}{g-g_*}+ d_1 +\mathcal{O}(g-g_*)  \right] \Gamma^{(n)}_R\,.
\end{align}
At leading order in perturbation theory, the solutions of the above equation is of the form
\begin{equation}\label{ASI2}
\Gamma^{(n)}_R \propto (g-g^*)^{d_0}e^{d_1(g-g^*)}\,,
\end{equation}
and its Borel transform is
\begin{equation}\label{ASI3}
\mathcal{B}\left(\Gamma^{(n)}_R\right) \propto (z/d_1)^{\frac{d_0-1}{2}} I_{d_0-1}(2\sqrt{d_1z})\,,
\end{equation}
where $I_n(z)$ is the modified Bessel function of the first kind. This function has a branch cut discontinuity in the complex plane from $z\in (-\infty,0]$ but is free of singularities in the positive and real axis. Nevertheless, when $g (g_*)$ is large enough, one still has to deal with the renormalons emerging from the skeleton diagram.
This contradiction may be solved by requiring the values of the coupling constants to lie inside the regions found in Ref.~\cite{Maiezza:2018pkk}~\footnote{This condition is in agreement with Ref.~\cite{Luty:2012ww}, where  it was concluded that the only asymptotic behaviors  described by perturbation theory are  scale invariant.}, on top of the usual condition of vanishing $\beta-$functions (criticality).

Our interpretation is that the renormalons should be understood as the manifestation in the perturbative expansion of the dynamically generated scale due to  \emph{dimensional transmutation}~\cite{PhysRevD.7.1888}. The dimensionful constant $\Lambda$ in Eqs.~\eqref{final} and~\eqref{nonlocal_CT} explicitly breaks scale invariance indeed.
The regions~\footnote{These regions are Reinhardt domains in the couplings space.} found in Ref.~\cite{Maiezza:2018pkk} are, as a matter of fact,  the domains of \emph{universality}~\cite{Wilson:1973jj} in the coupling's space. For coupling constant values beyond such regions,  the dimensional transmutation scale appears,  and this is by definition a non-universal feature.  Furthermore, within such regions the non-perturbative expressions can be resummed and obtained for example by means of the Meijer G-function algorithm recently proposed in Ref.~\cite{Mera:2018qte} and applied to 4D QFT in Ref.~\cite{Antipin:2018asc}. For instance, gauge theories with large number of fermions $N_f$ may be non-universal but critical~\cite{Antipin:2018asc} for some $N_f$.

\section{Conclusions}\label{conclusions}

The renormalons are a byproduct of the renormalization, which is purely a QFT feature and therefore they should not have a priori any semiclassical interpretation, although this quest has often been  tried in the literature. In this work, it is suggested that renormalons should be treated as the original divergences encountered in the early days of QFTs. The general approach underlying the whole paper is to start from a perturbatively renormalizable model, that provides the ``seed'' for going to the non-perturbative regime with the help of some recursive relations among different topologically disconnected sectors. We partially built-in the discussion based on the renormalon's notion, since their existence is a symptom of the failure of the perturbative renormalization program when the coupling is large enough. Our analysis may be regarded as a bridge between the perturbative and Wilsonian renormalizations, provided by the renormalons within the context of analyzable functions.
However, one should keep in mind that other ``bad series'' may likely emerge in the perturbative expansion, similar to the one for the renormalons. In any case, it is reasonable to assume that such series do not invalidate the conclusions of the present work.

For the sake of the argument, in the first part of the paper,  we have reviewed the state-of-the-art wisdom regarding the analytic structure of the Borel transform of the Green's functions. Afterwards,  we have shown that renormalons can be consistently resummed, leaving only one unknown constant to be determined \emph{a posteriori}. This unknown constant poses no difficulties, for it may be fixed by using a single boundary condition of phenomenological origin.

We have exploited the theory of the analyzable functions in the benchmark framework of the quark-antiquark potential with a large number of fermions.  This makes manifest the relevance of the whole approach and shows that the result of physical quantities depends upon one unknown,  non-perturbative constant,
as discussed around Eq.~\ref{final}, which is one of our central result. The resummed renormalons capture the expected behavior of the potential for large and small separations of the quark-antiquark.

Generalizing the perturbative renormalization approach used for the $\phi^4$ model, the effective Lagrangian that parameterizes the effects of the ultraviolet renormalons,  is the usual perturbatively renormalizable one augmented by some new counterterms. These new counterterms  are  non-local in space while  local in time  and make  the Lagrangian  non-renormalizable. We interpret the scale at which these non-local  operators emerge as a non-perturbative, non-universal dynamical scale,  not related to new degrees of freedom (or fields), but related instead  to the  dimensional transmutation energy scale.

We have studied the implications for perturbatively renormalizable models and provided some new insights on the analytic structure of the Borel transform of the Green's functions for scale-invariant theories. In particular, we suggest that renormalons should be understood as the manifestation in the perturbative expansion of the explicit breaking of the scale invariance due to the dimensional transmutation scale. Consequences on the asymptotic safety paradigm are also briefly discussed. We comment \emph{en passant} on the analogy between the usual safe QFT with coupling constant $g$ and asymptotically safe quantum gravity. The link is among  $g \leftrightarrow \varepsilon$, being $\varepsilon$ a small parameter for space-time dimensional expansion: $D=2+\varepsilon$. The perturbative expansion in powers of $g$ is the initial and the fundamental seed for the resurgent analysis in this work. A similar situation might happen for $\varepsilon$ in asymptotically safe quantum gravity~\cite{1977,Weinberg:1980gg}. Further calculations along this line is left for a future work.

\section*{Acknowledgement}

We thank Oleg Antipin for interesting discussions and for careful reading the manuscript, Daniele Dorigoni for clarifications on the resurgence of a Riccati's equation,
and Goran Senjanovi\'c for encouragement on this project.
AM was partially supported by the Croatian Science Foundation project number 4418. JV was funded by  Fondecyt project N. 3170154.

\appendix

\section{Highlights on resurgent functions}\label{app1}

For the sake of completeness, here we provide a pocket summary on the generalized resummation for analyzable functions, extracted from Ref.~\cite{costin1998,Borel-Ecalle}. For the simplest and logarithm free trasseries, the generalized Borel resummation can be summarized as in the following ~\cite{costin1998,Borel-Ecalle}:
\begin{enumerate}
\item
Given a formal divergent series denoted by  $y_0$,  one considers the associated formal transseries:
\begin{equation}
y(x) = y_0(x) + \sum_{k=1}^{\infty} C^k e^{-k\zeta/x}y_k(x)\,.
\end{equation}
where $C$ is arbitrary and complex, the poles in the positive real axis  of the Borel transformed function $\mathcal{B}(y_0)$ are  at $z_{pole}= \zeta ,2\zeta ,3\zeta ,.. $. It is a theorem for ODE that the singularities of $\mathcal{B}(y_k)$ (for $k\geq 1$) are also those  in $\mathcal{B}(y_0)$, together with a finite number of new ones\footnote{See theorem 5.67 of Ref.\cite{Borel-Ecalle} }. Notice that this is also the heart of the Bogomolny-Zinn-Justin prescription for instantons~\cite{Bogomolny:1980ur,ZinnJustin:1981dx}, where the imaginary ambiguities of $y_0$ are canceled against the imaginary ambiguities of the $y_k$. As we shall see,  the Costin prescription guarantees the reality of the resummed result, so that this cancellation mechanism is automatically taken into account. Finally, $y_0$ is the usual function that one finds in QFT using perturbation theory with the loop expansion, the other terms are invisible to perturbation theory around the origin.

\item
For each function $Y_k(z)$ $\forall k \in \mathbb{N}+ \{0\}$  one builds the functions $Y^{\pm}_k(z) \equiv Y_k(z\pm i \epsilon)$.

\item Resurgence:  once $Y_0$ is known to all orders  \emph{resurgence} is the property that allows the functions $Y_k$  to  be written in terms of $Y_0$  by the following operation:
\begin{equation} \label{eq:resurgence}
S_0^k Y_k=(Y_0^- - Y_0^{-k-1+}) \circ \tau_k, \quad \tau_k: z \mapsto z +k\zeta\,,
\end{equation}
where $S_0$ is a  non-perturbative  constant and
\begin{equation}\label{eq:analyticcontinuation}
Y_k^{-m+}  = Y_k^+ + \sum_{j=1}^m  {k+j \choose k} S_0^j  Y^+_{k+j}\circ \tau_{-j}\,.
\end{equation}
There is one Stokes constant associated with each singular direction containing the singularities $k\zeta _i$ for $i=,1,2,3,..,m$. In the present example and for simplicity,  we consider one singular direction with singularities in the real axis at $z_{pole}= \zeta ,2\zeta ,3\zeta ,...$.

Next  one can construct  the \emph{balanced average} associated to each $Y_k$
\begin{equation}\label{balanced:average}
Y_k^{bal} \equiv Y_k^+ + \sum_{n=1}^{\infty}2^{-n} (Y_k^- - Y_k^{-n-1+})\,.
\end{equation}
This definition guarantees that  when  $y_0$ is a real formal series, then $Y_k^{bal}$ is  also real on $\mathbb{R}-\{n\lambda\}$  and in this case the formula can be symmetrized by taking $1/2$ of the above expression plus $1/2$ of the same expression with $+$ and $-$ interchanged~\cite{costin1998,Borel-Ecalle}  (see proposition (5.77) and Eq.~(5.118) of Ref.~\cite{costin1998,Borel-Ecalle}).

\item Laplace Transform along a Stoke's direction: when the $Y_k(z)$ has poles in the positive real axis the Laplace transform is modified and  the summed result is given by
\begin{equation}
\mathcal{E}(y_k)= \mathcal{L \circ B}(y_k) = \mathcal{L}(Y_k) = \int_0^{\infty} Y_k^{bal} e^{-z/x}dz\,,
\end{equation}
such that the actual result of summing the original series $y_0(x)$ is obtained as
\begin{equation}
\mathcal{\sigma}(y_0(x))  \mapsto \mathcal{E}(y_0)(x) + \sum_{k=1}^{\infty} e^{-k\zeta /x} \mathcal{E}(y_k) (x)\,,
\end{equation}
where $\sigma$ denotes the generalized operation of the Borel resummation. When no poles are  present in the positive real axis the usual Borel procedure is recovered.
\end{enumerate}

The simplest example is the one of the Euler equation
\begin{equation}\label{EulerEq}
 -x^2y' +y=x\,.
\end{equation}
Borel transforming one gets a solution (in Borel space) as
\begin{equation}
\mathcal{B}(y_0) = \frac{1}{1-z}\,,
\end{equation}
with a pole in $z=1$. Applying directly the aforementioned four steps and defining $S_0\equiv i \frac{S}{2 \pi}$, one gets the complete solution of~\eqref{EulerEq}
\begin{equation}
y(x) \mapsto \sigma(y(x))= e^{-1/x}\text{Ei}\left(1/x\right) - \frac{4\pi^2C}{{S}}e^{-1/x}\,,
\end{equation}
which is manifestly real, i.e. with no imaginary ambiguity.  It is worth to comment that a simple non-linear modification to the Euler's equation leads to a Riccati's equation, whose solution has infinite singularities  similarly to the case of renormalons (i.e. singularities on the positive axis for some choice of the parameter), but with the difference that the poles are not simple~\cite{Dorigoni:2014hea}.


\bibliographystyle{jhep}
\bibliography{biblio}

\end{document}